# A comprehensive thermodynamic model for temperature change in *i*-caloric effects

A. M. G. Carvalho[1,2,3,*] and W. Imamura[1,4,5]

[1] *Departamento de Engenharia Mecânica, Universidade Estadual de Maringá, 87020-900, Maringá, PR, Brazil.*
[2] *Departamento de Engenharia Química, Universidade Federal de São Paulo, 09913-030, Diadema, SP, Brazil.*
[3] *Instituto de Física Armando Dias Tavares, Universidade do Estado do Rio de Janeiro, UERJ, Rua São Francisco Xavier, 524, 20550-013, Rio de Janeiro, RJ, Brazil.*
[4] *Departamento de Química, Universidade Estadual de Maringá, 87020-900, Maringá, PR, Brazil.*
[5] *Centro de Tecnologia, Universidade Federal de Alagoas, 57072-970, Maceió, AL, Brazil*

[*] *Corresponding author. E-mail: amgcarvalho@unifesp.br*

**ABSTRACT**

Solid-state cooling based on *i*-caloric effects may be an alternative to conventional vapor-compression refrigeration systems. The adiabatic temperature change ($\Delta T_S$) is one of the parameters that characterize the *i*-caloric effects, therefore it is important to obtain the correct $\Delta T_S$ values and, whenever possible, to correlate this parameter with thermodynamic and microscopic quantities. In this work, we propose a comprehensive thermodynamic model that allows us to determine the adiabatic temperature change from non-adiabatic measurements of temperature change induced by a field change. Our model fits efficiently temperature versus time and temperature change versus the inverse of the field change rate data for three different materials presenting different *i*-caloric effects. The results indicate the present model is a very useful and robust tool to obtain the correct $\Delta T_S$ values and to correlate $\Delta T_S$ with other thermodynamic quantities.

Solid-state cooling based on *i*-caloric effects may be the next-generation of refrigeration technologies and have received much attention in the last decades. *i*-caloric effect can be referred to the change in temperature or entropy, in reponse to an adiabatic or isothermal process, respectivelly, via application of external stimuli of field changes upon a given material. The change in temperature and entropy should be, at least, partially reversible, e.g., if temperature increases and entropy decreases when applying a field, temperature should descrease and entropy should increase when removing the field. Depending on the type of external field — such as mechanical, electric, and magnetic fields —, the *i*-caloric effect is called mechanocaloric, eletrocaloric and magnetocaloric. Mechanocaloric effect includes the particular cases elastocaloric effect (driven by uniaxial stress), barocaloric effect (driven by hydrostatic pressure) and twistocaloric (driven by pure torsion), besides more general cases.

The search for materials that present *i*-caloric effects large enough to be used in cooling technology has became a challenge. Large *i*-caloric effects can be found in intermetallics,[1,2,3,4,5,6,7] ceramics,[8] plastic crystals,[9] alkanes,[10,11] spin-crossover systems,[12,13,14] composites[15] etc. In general, large *i*-caloric effects appears in materiais around first- or second-order transitions, but polymers[16,17,18,19,20,21,22] can also exhibit large effects with or without transitions.

Adiabatic temperature change ($\Delta T_S$) and isothermal entropy change ($\Delta S_T$) are the main parameters that characterize the *i*-caloric effects. If one of these parameters is large, we say the *i*-caloric effect is large. For refrigeration, it is desirable that both parameters are large in certain conditions. Understanding the behavior of $\Delta T_S$ and $\Delta S_T$ as a function of temperature, applied field and other parameters is important to correlate the *i*-caloric effect with microscopic and thermodynamic quantities.

In this work, we focus our attention to the temperature change due to field change. Based on thermodynamic models applied to magnetocaloric effect[23] and elastocaloric effect,[24] we propose a comprehensive model to understand the temperature change behavior observed in *i*-caloric materials and to obtain further information:

$$\rho(i)\, c(i)\, \dot{T}(t) = -h(i)(T(t) - T_1) + \dot{W}_i(t) + \rho(i)\, \Delta s\, T(t)\, \dot{x}(t)\ , \qquad (1)$$

where *i* is the intensive variable that changes in time and provokes the corresponding *i*-caloric effect; *t* is the time; $\dot{T} \equiv dT/dt$ and *T* is the temperature; $T_1$ is the initial temperature; $\rho$ and c are the density and the specific heat of the material, respectively, which may depend on the variation of the intensive variable *i*; *h* is the volumetric heat transfer coefficient, $h(i) = h_0 \frac{A(i)}{V(i)}$, where $h_0$ is the heat transfer coefficient, *A* is the heat transfer surface area and *V* is the material volume; $\dot{W}_i \equiv dW_i/dt$ and $W_i$ is the work (per volume unit) done by intensive variable *i* on the material, not considering the latent heat due to a first-order transition; $\Delta s$ is the specific entropy variation at the transition presenting latent heat; $\dot{x} \equiv dx/dt$ and *x* is the mass fraction of one phase at a first-order transition.

As an initial approach, we consider processes where the external field variation, $\Delta i$, does not change significantly the volumetric heat transfer coefficient (*h*), the specific heat (*c*) and the material density ($\rho$), i.e., we keep *h*, *c* and $\rho$ fixed. Besides, $\Delta s$ does not depend on time. Then, integrating Eq. (1), we have

$$\int_{T_1}^{T_2} dT = -\frac{h}{\rho\, c} \int_{t_1}^{t_2} [T(t) - T_1]dt + \frac{1}{\rho\, c} \int_1^2 dW_i + \frac{\Delta s}{c} \int_{t_1}^{t_2} T(t)\, \dot{x}(t)dt\ . \qquad (2)$$

The term $\frac{\Delta s}{c} \int_{t_1}^{t_2} T(t)\, \dot{x}(t)dt$ represents the temperature change $\Delta T_{FOT}$ associated to a first-order transition. We also have $\int_1^2 dW_i = W_{12}$ and $\int_{t_1}^{t_2} [T(t) - T_1]dt = \overline{T - T_1}\, \Delta t$, with $\Delta t = t_2 - t_1$. $\overline{T - T_1}$ is the average temperature difference between the material and the reservoir during the external field variation. Considering the field changes in a constant rate *r*, then $\Delta t \sim \frac{1}{r}$ and $\overline{T - T_1} \cong$ $\frac{\Delta T}{2}$. Since $r \equiv \frac{di}{dt}$, we have $\Delta t = \frac{\Delta i}{di/dt}$, where $\Delta i$ is the field change between times $t_1$ and $t_2$. Therefore, $\overline{T - T_1} \Delta t = \frac{\Delta T}{2} \frac{\Delta i}{r}$. We can thus rewrite Eq. (2) as

$$\Delta T = -\frac{h}{2 \rho c} \frac{\Delta i}{r} \Delta T + \frac{1}{\rho c} W_{12} + \Delta T_{FOT} . \quad (3)$$

Imposing on Eq. (3) the adiabatic condition, $h = 0$, we have

$$\Delta T_S = \frac{1}{\rho c} W_{12} + \Delta T_{FOT} . \quad (4)$$

Subtracting Eq. (4) from Eq. (3), we get

$$\Delta T - \Delta T_S = -\frac{h}{2 \rho c} \frac{\Delta i}{r} \Delta T . \quad (5)$$

It is easy to see that

$$\Delta T = \frac{\Delta T_S}{1 + \frac{h}{2 \rho c} \frac{\Delta i}{r}} . \quad (6)$$

Comparing Eq. (6) with equation 6 from Ref. 24, we notice they are equivalent, with $\Phi \equiv \frac{h}{2 \rho c} \frac{\Delta i}{r}$. Here, it is important to point out three main differences between the present approach and that one reported in Ref. 24:

(a) The term $\frac{\Delta s}{c} \int_{t_1}^{t_2} T(t) \dot{x}(t) dt$ in Eq. (2) represents the temperature change associated to a first-order transition, not the adiabatic temperature change $\Delta T_S$;
(b) It is not possible to achieve Eq. (6) from Eq. (1) using the approach from Ref. 24;
(c) The present approach is simpler and more direct.

We can use Eq. (6) to analyze datasets of $\Delta T$ as a function of $r$ or $1/r$. Since we have a $\Delta i$ value for each rate $r$, we consider $\Delta i = \overline{\Delta i}$ and we find the best $\overline{\Delta i}$ to fit each $\Delta T$ vs. $1/r$ dataset in the present paper.

To analyze the behavior of the material's temperature as a function of time, we must solve Eq. (1) accordingly. Essentially, we solve Eq. (1) for two sequential processes: a field change process (from field $i_1$ and temperature $T_1$ to field $i_2$ and temperature $T_2$), hereinafter called *process $i_{1 \to 2}$*; and an isofield process (from $T_2$ to $T_1$, at constant field $i_2$), hereinafter called *process $T_{2 \to 1}$*.

For the process $i_{1 \to 2}$, the solution for Eq. (1) is

$$T(t) = \frac{1}{\mu(t)} \left\{ T_1 \mu(t_1) + \frac{1}{\rho c} \int_{t_1}^{t} \mu(t') [h T_1 + \dot{W}(t')] dt' \right\}, \quad (7)$$

where $\mu(t) = \text{Exp}\left[\frac{h}{\rho c}(t - t_1) - \frac{\Delta s}{c} \int_{t_1}^{t} \dot{x}(t') dt'\right]$.

For the process $T_{2 \to 1}$, $\dot{W}_i = 0$, and Eq. (1) becomes

$$\dot{T}(t) = -\frac{h}{\rho c}[T(t) - T_1] + \frac{\Delta s}{c}\dot{x}(t)\,T(t)\,. \tag{8}$$

Eq. (8) is solved for $t \geq t_2$, resulting in

$$T(t) = \frac{1}{\mu(t)}\left\{T_2\mu(t_2) + \frac{hT_1}{\rho c}\int_{t_2}^{t}\mu(t')dt'\right\}. \tag{9}$$

To obtain the adiabatic temperature change ($\Delta T_S$), we impose on Eq. (1) the adiabatic condition, $h = 0$. Thus, Eq. (1) becomes

$$\dot{T}(t) = \frac{\dot{W}(t)}{\rho c} + \frac{\Delta s}{c}\dot{x}(t)\,T(t)\,. \tag{10}$$

Eq. (10) is solved for $t_1 \leq t \leq t_2$, resulting in

$$T(t) = \frac{1}{\nu(t)}\left\{T_1\nu(t_1) + \frac{1}{\rho c}\int_{t_1}^{t}\nu(t')\dot{W}(t')dt'\right\}, \tag{11}$$

where $\nu(t) = \mathrm{Exp}\left[-\frac{\Delta s}{c}\int_{t_1}^{t}\dot{x}_{ph}(t')dt'\right]$. It is not difficult to see that

$$\Delta T_S = T(t_2) - T_1 = \frac{1}{\nu(t_2)}\left\{T_1\nu(t_1) + \frac{1}{\rho c}\int_{t_1}^{t_2}\nu(t)\dot{W}(t)dt\right\} - T_1\,. \tag{12}$$

To test our approaches, we apply the present model to different materials and different $i$-caloric effects. We use Eqs. (7), (9) and (12) to fit temperature vs. time data in order to obtain $\Delta T_S$. We also use Eq. (6) to fit the temperature change ($\Delta T$) vs. rate$^{-1}$ in order to obtain $\Delta T_S$. If our approaches are satisfactory, both $\Delta T_S$ values should be close.

Experimental data for magnetocaloric effect in metallic gadolinium[23] and the fits from the present model are shown in Fig. 1. The functions and parameters used in the model are listed in Table S1 (in the Supplementary Information). Temperature ($T$) vs. time ($t$) data and $\Delta T$ vs. $r^{-1}$ data were obtained at ~292 K, which is very close to the Curie temperature of Gd.

Firstly, we fit $T$ vs. $t$ data [Fig.1(a)] for $t \geq t_2$ using Eq. (9), which requires the function $\dot{x}(t)$ and the parameters $\Delta s$, $h$, $c$ and $\rho$. As magnetic transition of Gd does not present phase coexistence or latent heat, $\dot{x}(t) = 0$ and $\Delta s = 0$. In this case, we need three independent parameters to perform the fitting procedure. As we know $c$ and $\rho$ values for Gd,[25,26] we only have to find $h$. After that, we fit $T$ vs. $t$ data for $t_1 \leq t \leq t_2$, using Eq. (7), which requires the function $\dot{W}(t)$ and the parameters $h$, $c$ and $\rho$. Since we have $h$, $c$ and $\rho$, we only have to find an appropriate $\dot{W}(t)$, which is described in Table S1.

For $t_1 \leq t \leq t_2$, with $\mu_0\Delta H = 3$ T and $r = 0.01$ T s$^{-1}$, experimental $\Delta T$ is 5.6 K. In order to obtain the adiabatic temperature change ($\Delta T_S$), we use Eq. (12), which requires the functions and parameters previously obtained. Then, we get $\Delta T_S = 6.7$ K (20% higher than the measured $\Delta T$), indicating that the experimental process $i_{1\to 2}$ with a rate of 0.01 T s$^{-1}$ is far from the adiabatic condition.

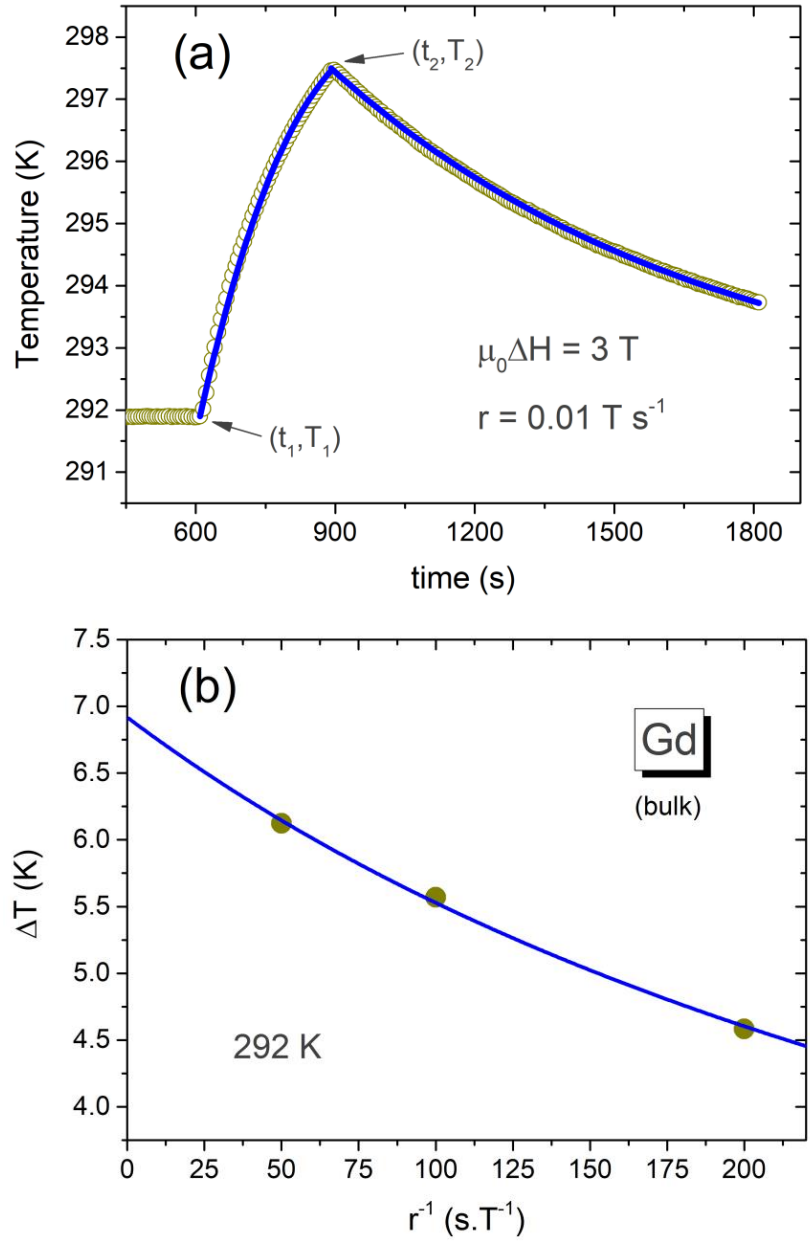

FIG. 1. Results for metallic gadolinium in bulk. (a) Temperature vs. time data, where temperature increases due to a positive magnetic field change (from 0 up to 3 T), with a rate of 0.01 T s$^{-1}$;[23] (b) Temperature change vs. rate$^{-1}$, where experimental data was obtained at 292 K.[23] Symbols represent experimental data and lines represent the fits from the model.

Magnetic-field-induced $\Delta T$ for Gd is shown in Fig. 1(b) as a function of the inverse of the rate of magnetic field change. The three experimental points[23] are fitted using Eq. (6). Since we already have $h$, $c$ and $\rho$ values, we have to find $\Delta i$ and $\Delta T_S$ values for the best fit. Here we get $\Delta T_S$

= 6.9 K. Comparing with $\Delta T_S$ obtained from $T$ vs. $t$ data, we note a difference of only 3%, which shows our approaches are valid in this case.

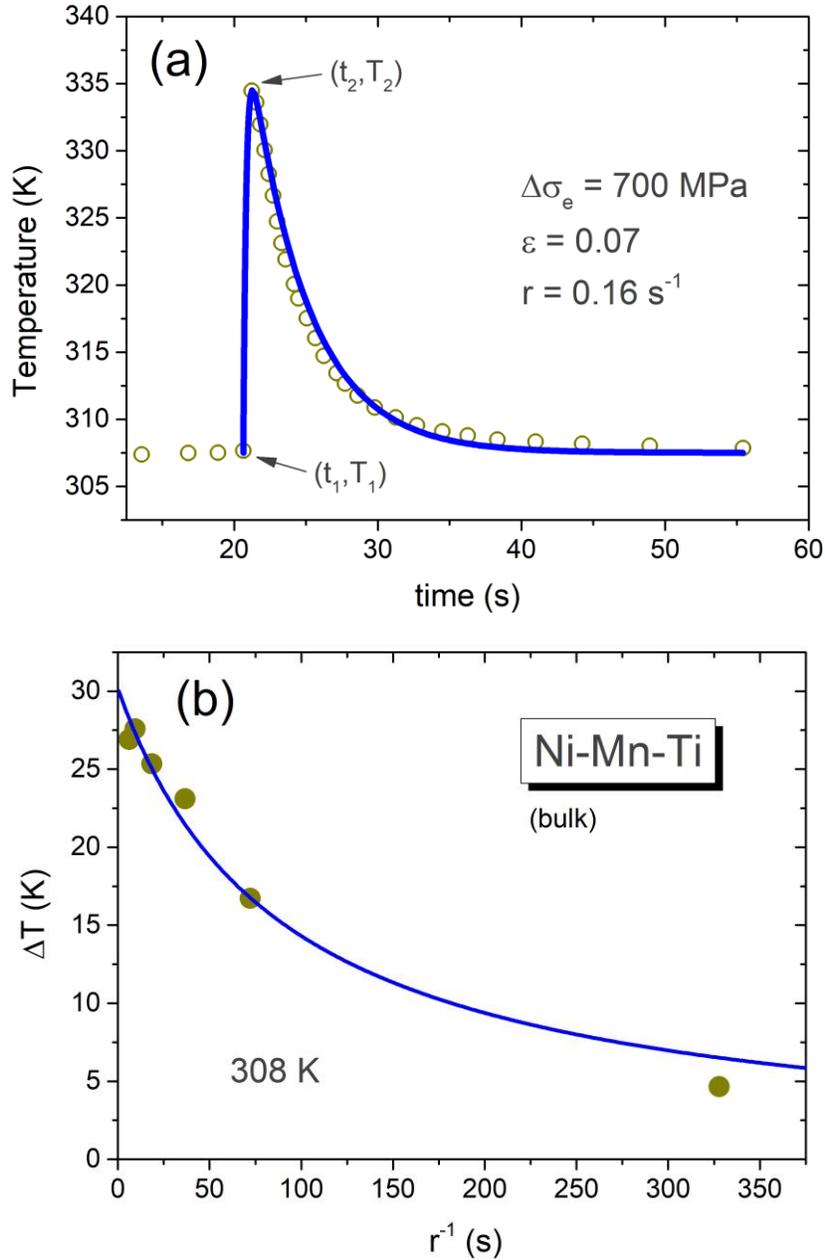

FIG. 2. Results for $(Ni_{50}Mn_{31.5}Ti_{18.5})_{99.8}B_{0.2}$ in bulk. (a) Temperature vs. time data, where temperature increases due to a positive compressive tension change of 700 MPa, resulting in a strain ($\varepsilon$) of 0.07, with a rate of 0.16 s$^{-1}$; (b) Temperature change vs. rate$^{-1}$, where experimental data was obtained at 308 K.[27] Symbols represent experimental data and lines represent the fits from the model.

Fig. 2 shows experimental data for compressive elastocaloric effect in $(Ni_{50}Mn_{31.5}Ti_{18.5})_{99.8}B_{0.2}$ (in bulk)[27] and the fits from the present model. The functions and parameters used in the model in this case are listed in Table S1 (in the Supplementary Information). $T$ vs. $t$ data and $\Delta T$ vs. $r^{-1}$ data were obtained at 308 K, which is above martensite-austenite transition temperature.[27]

As performed for Gd, we firstly fit $T$ vs. $t$ data for Ni-Mn-Ti alloy [Fig.2(a)] in the interval $t \geq t_2$, using Eq. (9), which requires the function $\dot{x}(t)$ and the parameters $\Delta s$, $h$, $c$ and $\rho$. Since we know $c$ and $\rho$ values for this material,[27,28] we have to determine $\dot{x}(t)$, $\Delta s$ and $h$ for the best fit. As we have three independent parameters to find, we optimized this fit in conjunction with the fit of the interval $t_1 \leq t \leq t_2$, which also requires the function $\dot{W}(t)$. According to Ref. 27, $\Delta s = 76$ J kg$^{-1}$ K$^{-1}$ for $(Ni_{50}Mn_{31.5}Ti_{18.5})_{99.8}B_{0.2}$. Using this value, we were not able to fit both intervals satisfactorily. For the fit shown in Fig. 2(a), $\Delta s = 46$ J kg$^{-1}$ K$^{-1}$; besides, $\dot{W}(t) = 0$ [and $W(t) = 0$], which is consistent with the hypothesis that, in this case, the temperature change is entirely due to the structural phase transition. Considering that $\Delta s$ previously reported[27] is correct, the divergence between the reported value and the theoretical one suggests two scenarios (that may coexist): (1) the specific heat of austenite phase is significantly different from specific heat of strain-induced martensite phase around the structural transition, affecting theoretical $\Delta s$ value [since the ratio $\Delta s/c$ appears in both Eqs. (7) and (9)]; (2) the strain-induced transition is not complete, then $\Delta s$ from $\Delta T$ experiment is lower than $\Delta s$ from differential scanning calorimetry (DSC).

For $t_1 \leq t \leq t_2$ [with $\Delta \sigma_e = 700$ MPa, strain ($\varepsilon$) of 0.07 and rate ($r$) of 0.16 s$^{-1}$], experimental $\Delta T$ is 26.9 K. In order to obtain the adiabatic temperature change ($\Delta T_S$), we use Eq. (12), which requires the functions and parameters previously obtained in the fitting procedure. Then, we get $\Delta T_S = 29.9$ K (11% higher than the measured $\Delta T$), indicating that the experimental process $i_{1 \to 2}$ with the rate of 0.16 s$^{-1}$ is not close to the adiabatic condition. Here we see that only a very fast change in the intensive variable $i$ is not enough to stablish a quasi-adiabatic process. In this case, the time for the process $i_{1 \to 2}$ is very small (less than 1 s), but the theoretical volumetric heat transfer coefficient is enormous ($h = 8.2 \times 10^5$ W m$^{-3}$ K$^{-1}$).

Strain-induced $\Delta T$ for Ni-Mn-Ti alloy is shown in Fig. 2(b) as a function of the inverse of the strain rate. Several experimental points[27] are fitted using Eq. (6). Since we already have $h$, $c$ and $\rho$ values, we have to find $\Delta i$ and $\Delta T_S$ values for the best fit. Here we get $\Delta T_S = 30.2$ K. Comparing with $\Delta T_S$ obtained from $T$ vs. $t$ data, we note a difference of only 1%, which shows our approaches are also valid in this case.

Fig. 3 shows experimental data for tractive elastocaloric effect in $Ni_{50.4}Ti_{49.6}$ (20-μm-thick films)[29] and the fits from the present model. The functions and parameters used in the model in this case are listed in Table S1 (in the Supplementary Information). $T$ vs. $t$ data and $\Delta T$ vs. $r^{-1}$ data were obtained at room temperature, near martensite-austenite transition temperature[29].

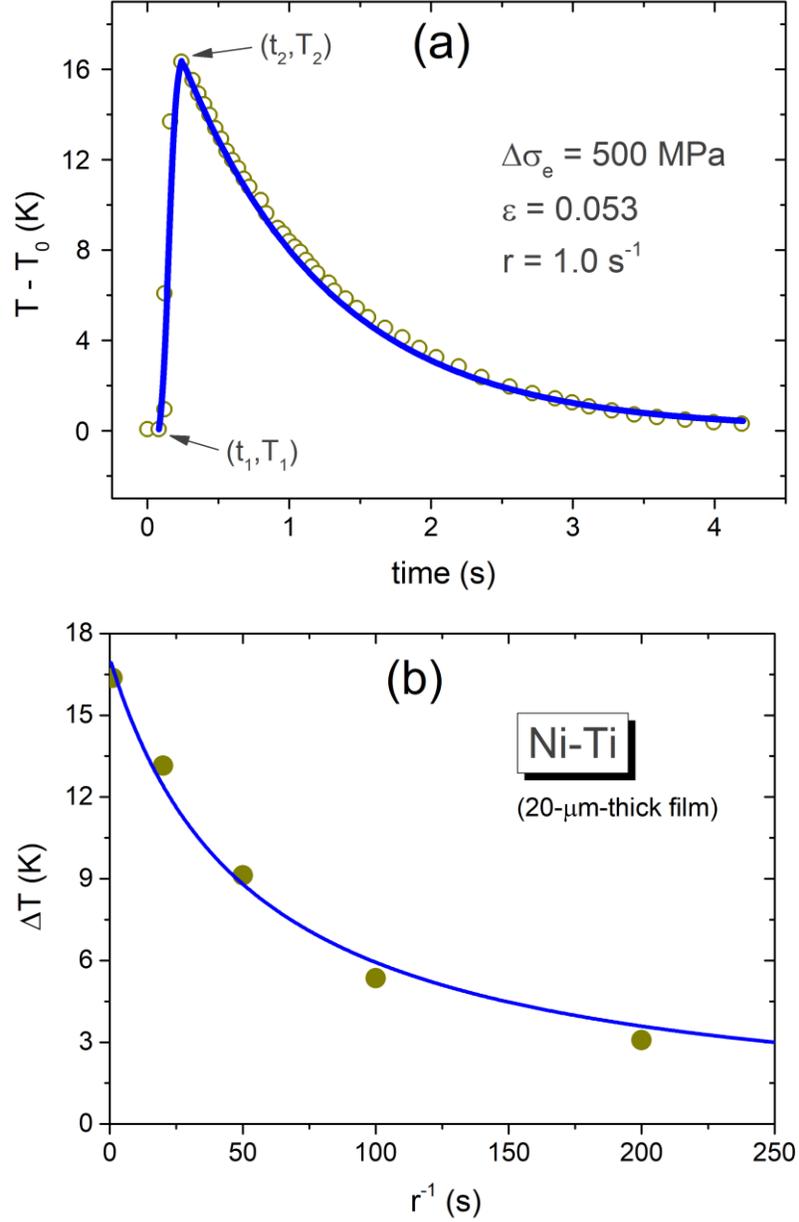

FIG. 3. Results for 20-μm-thick films of $Ni_{50.4}Ti_{49.6}$. (a) Temperature vs. time data, where temperature increases from room temperature ($T_0$) due to a positive tractive tension change of 500 MPa, resulting in a strain of 0.053, with a rate of 1.0 s$^{-1}$; (b) Temperature change vs. rate$^{-1}$, where experimental data was obtained at room temperature.[29] Symbols represent experimental data and lines represent the fits from the model.

As performed for the previous materials, we firstly fit $T$ vs. $t$ data for Ni-Ti film [Fig. 3(a)] in the interval $t \geq t_2$, using Eq. (9), which requires the function $\dot{x}(t)$ and the parameters $\Delta s$, $h$, $c$ and $\rho$. Since we know $c$ and $\rho$ values for this material,[29] we have to determine $\dot{x}(t)$, $\Delta s$ and $h$ for the best fit. As in the case of Ni-Mn-Ti, we have three independent parameters to find, then we optimized this fit in conjunction with the fit of the interval $t_1 \leq t \leq t_2$, which also requires the function $\dot{W}(t)$. For Ni-Ti, we get $\dot{W}(t) \neq 0$ and $\Delta s = 23$ J kg$^{-1}$ K$^{-1}$. The reported latent heat for Ni-Ti film, obtained from DSC,[29] is 20 kJ kg$^{-1}$, resulting in $\Delta s \cong 74$ J kg$^{-1}$ K$^{-1}$. Interestingly, it is also mentioned that from $\Delta T$ experiment the estimated latent heat is much lower (7.2 kJ kg$^{-1}$), indicating that the material only undergoes part of the phase transformation during load cycling, according to Ref. 29. The latent heat of 7.2 kJ kg$^{-1}$ results in $\Delta s = 24$ J kg$^{-1}$ K$^{-1}$ (considering 300 K as the reference temperature), which is very close to the theoretical $\Delta s$ value (23 J kg$^{-1}$ K$^{-1}$) used in the fitting procedure.

For $t_1 \leq t \leq t_2$ (with $\Delta\sigma_e = 500$ MPa, $\varepsilon = 0.053$ and $r = 1.0$ s$^{-1}$), experimental $\Delta T$ is 16.4 K. In order to obtain the adiabatic temperature change ($\Delta T_S$), we use Eq. (12), which requires the functions and parameters previously obtained in the fitting procedure. Then, we get $\Delta T_S = 17.7$ K (8% higher than the measured $\Delta T$), indicating that the experimental process $i_{1\rightarrow 2}$ for Ni-Ti, with the rate of 1.0 s$^{-1}$, is closer to the adiabatic condition than Ni-Mn-Ti (11% higher than the measured $\Delta T$), with the rate of 0.16 s$^{-1}$. The strain rate for Ni-Ti is 6.3 times larger than the rate for Ni-Mn-Ti, and the theoretical volumetric heat transfer coefficient, $h$, for Ni-Ti is $2.8\times 10^6$ W m$^{-3}$ K$^{-1}$, 3.4 times larger than the theoretical $h$ for Ni-Mn-Ti. From the definition of the volumetric heat transfer coefficient ($h = h_0 A/V$, where $A$ heat transfer surface area and $V$ is the material volume), this result is consistent, since Ni-Mn-Ti is a bulk sample while Ni-Ti is a 20-μm-thick film (much higher $A/V$ ratio).

Strain-induced $\Delta T$ for Ni-Ti is shown in Fig. 3(b) as a function of the inverse of the strain rate. Five experimental points[29] are fitted using Eq. (6). Since we already have $h$, $c$ and $\rho$ values, we have to find $\Delta i$ and $\Delta T_S$ values for the best fit. Here we get $\Delta T_S = 17.1$ K. Comparing with $\Delta T_S$ obtained from $T$ vs. $t$ data, we note a difference of only 3%, which shows our approaches are also valid in this case.

For Ni-Ti, $\Delta T_S$ obtained from $T$ vs. $t$ data has higher values for much lower strain rates. We found 19.2 K and 20.1 K for the strain rates of 0.05 s$^{-1}$ and 0.02 s$^{-1}$, respectively. During $\Delta T$ experiments, Ni-Ti films are stretched and $A/V$ ratio may increase significantly, i.e., $h$ may increase significantly. Eqs. (7) and (9), used to fit $T$ vs. $t$ data, were derived with the condition of constant $h$. Therefore, even with reasonable fits, it is expected that these equations may yield overestimated (or underestimated) $\Delta T_S$ values when $h$ changes significantly during the process $i_{1\rightarrow 2}$. Interestingly, if the process $i_{1\rightarrow 2}$ is not far from adiabatic condition (as is the case for Ni-Ti with strain rate of 1.0 s$^{-1}$), the fits and $\Delta T_S$ obtained are highly satisfactory. It is also interesting that $\Delta T$ vs. $r^{-1}$ data is satisfactorily fitted as well, even Eq. (6) being derived using the same consideration of constant $h$. In this case, the free parameter $\Delta i$ in Eq. (6) seems to compensate the variation of the volumetric heat transfer coefficient.

As a conclusion, the thermodynamic model proposed in this work allows us to determine the adiabatic temperature change ($\Delta T_S$) from non-adiabatic measurements of $\Delta T$ through two different approaches: (a) from temperature change vs. rate$^{-1}$ data and using eq. (6); (b) from temperature vs. time data and using eqs. (7) and (9). Our model fits efficiently temperature vs. time and temperature change vs. rate$^{-1}$ data for three different materials presenting different *i*-caloric effects: magnetocaloric effect in metallic gadolinium in bulk, tractive elastocaloric effect in $(Ni_{50}Mn_{31.5}Ti_{18.5})_{99.8}B_{0.2}$ in bulk and compressive elastocaloric effect in films of $Ni_{50.4}Ti_{49.6}$. In all examples presented and detailed in this paper, $\Delta T_S$ values obtained from both approaches are very close, showing both approaches and, consequently, our model are valid.

We noticed $\Delta T_S$ for Ni-Ti alloy obtained using the second approach (from *T* vs. *t* data) has higher values for much lower strain rates (not shown). This is due to the fact that, during $\Delta T$ experiments, Ni-Ti films are stretched and $h$ may increase significantly, since $h = h_0\, A/V$ and the ratio $A/V$ may increase significantly. Therefore, to use this approach, it is important to keep $h$ nearly constant during the experiments, which may be an issue when stretching films (in general) and elastomers.

Analyzing eq. (9), it is not difficult to see that if there is not latent heat and phase coexistence [$\Delta s = 0$ and $\dot{x}(t) = 0$] and we know two of the parameters $h$, $c$ and $\rho$, we may find the third one by fitting the curve referent to the process $T_{2 \to 1}$. For materials that do not present first-order transition, this should be valid in all temperatures and applied fields. For materials that present first-order transitions, this should be valid for temperature and applied field intervals off the region of phase coexistence.

In summary, the virtues of the present model indicate that it is a very useful and robust tool to obtain the correct $\Delta T_S$ values and to correlate $\Delta T_S$ with other thermodynamic quantities. Furthermore, this model is possibly valid for any *i*-caloric effect.

This work was supported by Conselho Nacional de Desenvolvimento Científico e Tecnológico – CNPq (Proc. 163391/2020-3).

**Supplementary Information**

# A comprehensive thermodynamic model for temperature change in *i*-caloric effects

A. M. G. Carvalho[1,2,3,*] and W. Imamura[1,4,5]

[1] *Departamento de Engenharia Mecânica, Universidade Estadual de Maringá, 87020-900, Maringá, PR, Brazil.*
[2] *Departamento de Engenharia Química, Universidade Federal de São Paulo, 09913-030, Diadema, SP, Brazil.*
[3] *Instituto de Física Armando Dias Tavares, Universidade do Estado do Rio de Janeiro, UERJ, Rua São Francisco Xavier, 524, 20550-013, Rio de Janeiro, RJ, Brazil.*
[4] *Departamento de Química, Universidade Estadual de Maringá, 87020-900, Maringá, PR, Brazil.*
[5] *Centro de Tecnologia, Universidade Federal de Alagoas, 57072-970, Maceió, AL, Brazil*

[*] *Corresponding author. E-mail: amgcarvalho@unifesp.br*

Table S1: Functions and parameters used to fit temperature vs. time data (figures 1a, 2a and 3a in the main text) and temperature change vs. rate$^{-1}$ data (figures 1b, 2b and 3b in the main text).

| | **Gd** | **Ni-Mn-Ti** | **Ni-Ti** |
|---|---|---|---|
| $\dot{W}(t)$ (W m$^{-3}$) | $M_H r\{1 - \tanh[ar(t - t_1)]\}$ [a] <br> $M_H = 5{,}95 \times 10^6$ J m$^{-3}$ T$^{-1}$ <br> $a = 0.22$ T$^{-1}$ <br> $r = 0.01$ T s$^{-1}$ | 0 [b] | $\dfrac{w}{t_w} e^{\frac{t-t_1}{t_w}}$ [c] <br> $w = 7.50 \times 10^6$ J m$^{-3}$ <br> $t_w = 0.22$ s |
| $\dot{x}(t)$ (s$^{-1}$) | 0 [d] | $\dfrac{1}{t_x} e^{-\frac{t-t_1}{t_x}}$ [e] <br> $t_x = 0.2$ s | $\dfrac{0.39894}{t_x} e^{-\frac{(t-t_1-t_c)^2}{2t_x^2}}$ [f] <br> $t_c = 0.07$ s <br> $t_x = 0.04$ s |
| $\Delta s$ (J kg$^{-1}$ K$^{-1}$) | 0 | 46 | 23 |
| $\rho$ (kg m$^{-3}$) | 7900 [g] | 7040 [h] | 6500 [i] |
| $c$ (J kg$^{-1}$ K$^{-1}$) | 250 [j] | 470 [k] | 450 [i] |
| $h$ (W m$^{-3}$ K$^{-1}$) | 2410 | $820 \times 10^3$ | $2.80 \times 10^6$ |
| $\Delta i$ | 4.1 T | 0.09 | 0.04 |

[a] $W(t) = M_H \left\{ H - \dfrac{1}{a}\ln[\cosh(aH)] \right\}$, where $H = r(t - t_1)$ for a constant rate of magnetic field change. [b] $W(t) = 0$. [c] $W(t) = we^{\frac{t-t_1}{t_w}}$. [d] $x(t)$ is constant. [e] $x(t) = -e^{-\frac{t-t_1}{t_x}}$. [f] $x(t) = \dfrac{1}{2}\left[1 + \text{Erf}\left(\dfrac{t-t_1-t_c}{\sqrt{2}t_x}\right)\right]$. [g] From Ref. 1. [h] From Ref. 2. [i] From Ref. 3. [j] From Ref. 4. [k] From Ref. 5.

**References for Supplementary Information**